\documentstyle[preprint,aps,pra]{revtex}

\begin{document}
\draft

\title{Ab~initio calculation of the KRb dipole moments}

\author{S. Kotochigova, P. S. Julienne, and E. Tiesinga}

\address{National Institute of Standards and Technology, 100 Bureau Drive,
stop 8423, Gaithersburg, Maryland 20899.}

\maketitle

\begin{abstract}
The relativistic configuration interaction valence bond method has been
used to calculate permanent and transition electric dipole moments of
the KRb heteronuclear molecule as a function of internuclear separation.
The permanent dipole moment of the ground state $X^1\Sigma^+$ potential
is found to be 0.30(2) $ea_0$ at the equilibrium internuclear separation with
excess negative charge on the potassium atom.
For the $a^3\Sigma^+$ potential the dipole moment is an order of
magnitude smaller (1 $ea_0=8.47835\,10^{-30}$ Cm) 
In addition,
we calculate transition dipole moments between the two ground-state
and excited-state potentials that dissociate to the
K(4s)+Rb(5p) limits.  Using this data we propose a way to produce
singlet $X^1\Sigma^+$ KRb molecules by a two-photon Raman process
starting from an ultracold mixture of doubly spin-polarized ground state
K and Rb atoms.  This Raman process is only allowed due to relativistic
spin-orbit couplings and the absence of gerade/ungerade selection rules
in heteronuclear dimers.  \end{abstract}



In heteronuclear diatomic molecules the electrons can be distributed
between the two nuclei unequally. As a result, there can be a small
excess of negative charge near the more electronegative atom and thus
heteronuclear dimers can have a permanent dipole moment.  In this
study we calculate the permanent dipole moments of the $^{1,3}\Sigma^+$
electronic states of the KRb ground configuration, which dissociate to the
K(4s)+Rb(5s) limit, and transition dipole moments between these states and
excited states, which dissociate to K(4s)+Rb(5p) limits. Simple analysis
of the ionization energies of K and Rb atoms, 4.339 eV and 4.176 eV,
respectively, shows that the K atom is most likely to have an excess of
a negative charge when the two atoms interact.

There are several applications of dipolar molecules in the physics
of ultra-cold molecular gases. The long-range interaction between two
dipolar molecules is governed by a electric dipole-dipole interaction
and can, for example, significantly modify the many-body dynamics
of trapped ultra-cold molecular Bose Einstein Condensates (BEC's)
\cite{You,Goral,Santos,Goral01,Meystre,Giovanazzi}. The atom-atom
interactions in current atomic BEC's are spherically symmetric
in origin.  Following the proposal for homonuclear molecules of
Ref.~\cite{Jaksch}, dipolar molecules might also be formed in an
optical lattice\cite{Damski,Moore} An ultra-cold atom of each of the
atomic species is held in an individual lattice site.  A molecule in the
electronic ground state can then be formed under the influence of laser
light by inducing a two-photon transition.  For any of the proposed
applications, knowledge of the permanent or transition dipole moments
is essential.

The dipole moments have never been determined for KRb states of the ground
configuration either theoretically or experimentally. We calculate the
dipole moment of the singlet $X^1\Sigma^+$ and triplet $a^3\Sigma^+$
states of KRb as a function of internuclear separation $R$ using an
{\it ab~initio} configuration interaction valence bond (VB) method
\cite{Kotochigova}. Both relativistic and nonrelativistic calculations
are presented, which allows us to determine the influence of relativistic
effects on the dipole moment. The use of configuration interaction (CI)
in the VB method is essential due to the role of correlation in the
formation of the molecular bond and nonzero dipole moment.  The main
contribution to the dipole moment of KRb is not due to the ground state
atomic configuration, even though it predominantly determines the energy,
but is due to the population of excited valence and virtual orbitals.

The accuracy of the calculation of the $X^{1}\Sigma^+$ wave
function is tested by comparing our {\it ab~initio} potential with
existing experimental \cite{Ross,Okada,Kasahara,Amiot} and theoretical
\cite{Yiannopoulou,Leininger,Park,Rousseau} singlet potentials.  We also
present the $a^{3}\Sigma^+$ potential, compare it with the
calculations of Refs.~\cite{Park,Rousseau}, and indicate the sensitivity
of this shallow potential to the computational method.

Furthermore, we analyze the possibility of the creation of dipolar KRb
molecules in an optical lattice \cite{Damski}. In this Paper we calculate
the transition electric dipole moments between the $^{1,3}\Sigma^+$
states of the ground configuration and relativistic $\Omega=0^\pm,1$
components of excited $^{1,3}\Sigma^+$ and $^{1,3}\Pi$ states dissociating
to the K(4s)+Rb(5p) limits, where $\Omega$ is the projection of the
total electronic angular momentum of the two atoms on the internuclear
axis. This data provides us information about the most efficient scheme of
forming KRb molecules in the singlet $X^{1}\Sigma^+$ ground state from a
K(4s)+Rb(5s) collision on the triplet $a^3\Sigma^+$ potential. Moreover,
we present our {\it ab~initio} potentials for the excited states which
can be used to identify the complex behavior of the transition dipole
moment as function of $R$.  Large-scale theoretical studies of the excited
potentials of KRb were previously performed in Refs.~\cite{Park,Rousseau}.
It was shown in Refs.~\cite{Bussery,Patil,Wang,Marinescu} that the
dispersion interaction of excited KRb states is considerably stronger than
in any other heteronuclear alkali dimer and that the KRb system is a good
candidate for photoassociation experiments. Transition dipole moments to
more-highly excited KRb states have been calculated in Ref.~\cite{Park}.

\section{Theory}

The electronic potentials and dipole moments are calculated with the
configuration-interaction (CI) valence bond method.  Both nonrelativistic
and relativistic implementations of the method are used.  The basic
idea behind the valence-bond method is that the electronic molecular
wavefunction $|\Psi_{AB}\rangle$ is constructed from wavefunctions,
which are products of wavefunctions that describe the constituent atoms
$A$ and $B$. In essence, the molecular wavefunction is given by
\begin{equation}
|\Psi_{AB}\rangle = \sum_{\alpha}C_{\alpha}|det_{\alpha}^{AB}\rangle \, \label{psi}
\end{equation}
where each $|det^{AB}\rangle$ is an antisymmetrized ($\hat A$) product of 
two atomic Slater determinants 
\begin{equation}
|det_{\alpha}^{AB}\rangle = {\hat A}(|det_{\alpha}^{A}\rangle \cdot 
                                      |det_{\alpha}^{B}\rangle ) \,.
\end{equation}
The atomic Slater determinants $|det_{\alpha}^{A}\rangle$ and
$|det_{\alpha}^{B}\rangle$ are constructed from one-electron functions
centered on the nucleus of atom $A$ and $B$, respectively.  The variational
CI coefficients $C_\alpha$ in Eq.~(\ref{psi}) are obtained by solving
a generalized eigenvalue matrix problem, since in our approach the
$|det_{\alpha}^{AB}\rangle$ are not orthogonal.

The permanent molecular dipole moment of the molecular wavefunction
$|\Psi_{AB}\rangle$ is calculated from
\begin{equation}
\vec{\mu} = \sum_{k=A,B}Z_k\vec{R_k} - 
       e\langle\Psi_{AB}|\sum_{i=1}^N\vec{r_i}|\Psi_{AB}\rangle
\\ \label{dipole}
\end{equation}
where $N$ is the total number of electrons in the molecule and $e$
is the electron charge. The first term of Eq.~\ref{dipole} depends
on the position $\vec{R}_k$ and charge $Z_k$ of nuclei $k=A$ and $B$ and
is zero by choosing an appropriate coordinate origin.
The second term of Eq.~\ref{dipole} depends on the electronic molecular
wavefunction and the electron positions $\vec{r}_i$. 
In the valence-bond method
\begin{equation}
\langle\Psi_{AB}|\sum_{i=1}^N \vec{r_i}|\Psi_{AB}\rangle 
            = N\sum_{\kappa} \int d\vec{r}\, \vec{r}\rho^{AB}(\vec{x},\vec{x}) \,,
\end{equation}
where $\vec{x}=(\vec{r},\kappa)$ denotes both the single-electron coordinate
$r$ and the variable $\kappa$. For a nonrelativistic or relativistic
calculation $\kappa$ has two or four values, respectively.  The  
single-electron density matrix is
\begin{equation}
\rho^{AB}(\vec{x},\vec{x}\,') = 
     \sum_{\alpha,\beta}C^*_\alpha C_\beta\rho^{\alpha,\beta}(\vec{x},\vec{x}\,')
  \,,
\end{equation}
where the transition density matrix between determinants
$|det^{AB}_{\alpha}\rangle$ and $|det^{AB}_{\beta}\rangle$ is defined by
\begin{eqnarray}
\lefteqn{\rho^{\alpha,\beta}(\vec{x},\vec{x}\,') \,=} 
\label{ro1} \\
&& 
 (D_{\alpha\alpha} D_{\beta\beta})^{-1/2} \cdot D_{\alpha\beta}
\sum_{i,j}^{N} (S^{-1})_{i,j}^{\alpha,\beta} \cdot \psi^*_i(\vec{x}) \cdot
\varphi_j(\vec{x}\,') \,,  \nonumber
\end{eqnarray}
$D_{\alpha\beta}$ is the
determinant of the $N \times N$ overlap matrix $S_{i,j}^{\alpha,\beta}$
= $\langle\psi_i|\varphi_j\rangle$, and the $\psi_i(\vec{x})=\langle
\vec{x}|\psi_i\rangle$ and $\varphi_i(\vec{x})=\langle
\vec{x}|\varphi_j\rangle$ denote the single-electron functions used
in constructing the basis functions $|det^{AB}_{\alpha}\rangle$
and $|det^{AB}_{\beta}\rangle$, respectively. Moreover, $D_{\alpha\alpha}$
and $D_{\beta\beta}$ are the determinants of the matrices 
$S_{i,j}^{\alpha,\alpha}=\langle\psi_i|\psi_j\rangle$
and $S_{i,j}^{\beta,\beta}=\langle\varphi_i|\varphi_j\rangle$.
The one-electron wavefunctions $|\psi_i\rangle$ and $|\psi_j\rangle$
are obtained by self-consistently solving Hartree-Fock and Sturmian
equations for a nonrelativistic calculation and Dirac-Fock and Sturmian equations
for a relativistic calculation \cite{Kotochigova}.
Transition dipole moments can be derived in a similar fashion by
assuming different initial and final states in Eq.~\ref{dipole}.

\section{Ground state dipole moments} \label{gs}

The atomic determinants for the KRb dimer are constructed from
single-electron Hartree-Fock or Dirac-Fock functions/orbitals for
electrons in closed and valence shells.  Sturmian wavefunctions
complement the basis functions and are used to describe virtual
orbitals. One-electron functions are characterized by a main quantum
number $n=1,2,\dots$ and orbital angular momentum $l=s,p,\dots$ for
both nonrelativistic and relativistic calculations. In addition for
relativistic calculations the one-electron orbital is labeled by the
total electron spin $j$.

The closed shells $1s^2~ 2s^2~ 2p^6~ 3s^2$ of K and $1s^2~
2s^2~ 2p^6~ 3s^2~ 3p^6~ 3d^{10}~ 4s^2$ of Rb form the core of
the heteronuclear molecule. The superscript denotes the number
of electrons in shell $nl$. For relativistic calculations $2p^6$
is short for $2p_{1/2}^22p_{3/2}^4$ etcetera. In our calculations
excitations from these closed shells to valence and virtual orbitals
are not included, i.e. all atomic determinants in the molecular basis
contain the same number of electrons in these orbitals.  Single electron
excitations from the closed $3p^6$ shell of K and $4p^6$ shell of Rb are
allowed and introduce core-valence correlations in the CI.  The $4s$,
$4p$, and $3d$ valence orbitals of K and $5s$, $5p$, and $4d$ orbitals
of Rb are allowed to contain at most two electrons. In addition, for the
nonrelativistic calculation we allow at most two electrons in the $4d$,
$5p$, and $6p$ virtual orbitals of potassium and the $5d$, $6p$, and
$7p$ virtual orbitals of rubidium.  For the relativistic calculation
computational limitations restricted us to the $4d$ and $5p$ virtual
orbitals of K and $5d$ and $6p$ virtual orbitals of Rb. Both covalent
and ionic configurations are constructed.  The CI expansions with these
basis sets have 493 nonrelativistic and 1459 relativistic configurations.

\begin{figure} \vspace*{11cm}
\includegraphics{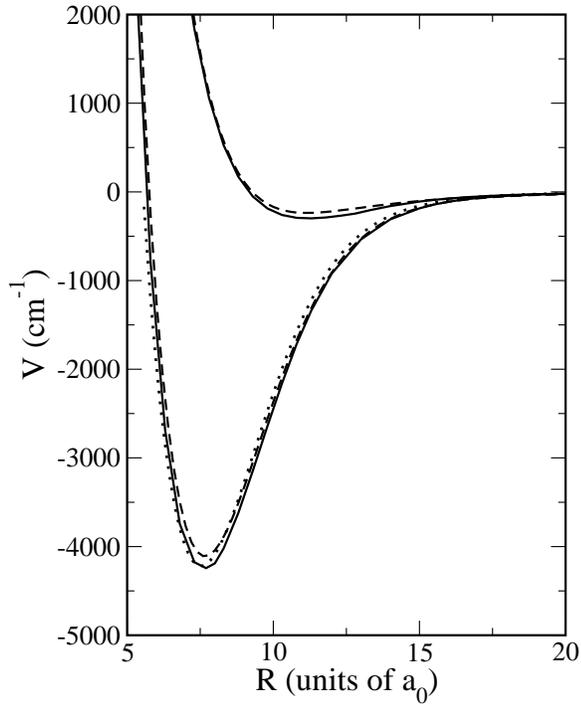}
\caption{ The ground state $X^{1}\Sigma^+$ and $a^{3}\Sigma^+$ potentials
of KRb as a function of internuclear
separation. Solid curves are obtained in the present study, dashed lines
are the results of Ref.~{\protect\cite{Rousseau}}, and the dotted line is the
RKR potential of Ref.~{\protect\cite{Amiot}}. (1 $a_0$=0.0529 nm)}
\label{potentials} \end{figure}

The ground state potentials have equilibrium distances $R_e$
and dissociation energies $D_e$ that agree with experimental
RKR values \cite{Kasahara,Amiot} and other theoretical
potentials. Figure~\ref{potentials} shows the ground state $X^{1}\Sigma^+$
and $a^3\Sigma^+$ potentials of the KRb dimer as a function of the
internuclear separation $R$. The solid curves of Fig.~\ref{potentials}
describe results of our relativistic ($\Omega$ = 0) {\it ab~initio}
calculation whereas the dashed curves are the theoretical data of
Ref.~\cite{Rousseau}. The dotted line of Fig.~\ref{potentials} shows
the singlet RKR potential of Ref.~\cite{Amiot}.

For the $X^{1}\Sigma^+$ state the difference between the two theoretical
calculations shown in Fig.~\ref{potentials} is about 3\% at $R_e$.  A 20\%
difference exists at $R_e$ of the $a^3\Sigma^+$ potential. Our estimate
of the $C_6/R^6$ Van-der-Waals coefficient for the $X^{1}\Sigma^+$
potential, obtained by fitting to the dispersion potential $C_6/R^6 +
C_8/R^8 + C_{10}/R^{10}$, agrees to 2-3\% with the value of Derevianko
et. al.\cite{Derevianko}.  The experimentally determined dissociation
energy for the $X^{1}\Sigma^+$ potential \cite{Kasahara,Amiot} is 26
cm$^{-1}$ smaller than ours.

We performed calculations of the $X^{1}\Sigma^+$ and $a^3\Sigma^+$ state
potentials and their permanent dipole moments in both nonrelativistic and
relativistic approximation. For the same set of one-electron orbitals
the influence of relativistic effects on the potential energy is very
small whereas its influence on the dipole moments is large. At $R_e$ a
relativistic calculation leads to a 30\% increase in the absolute value
of the dipole moment for the $X^1\Sigma^+$ state and a 50\% decrease for
the $a^3\Sigma^+$ state.  Moreover, we find large correlation effects in
the electric dipole moment in both nonrelativistic and relativistic
calculations. The implementation of core polarization, for instance,
leads a 50\% of reduction in the absolute value of the dipole moment at
the bottom of the singlet potential.

\begin{figure} \vspace*{12.5cm} \includegraphics{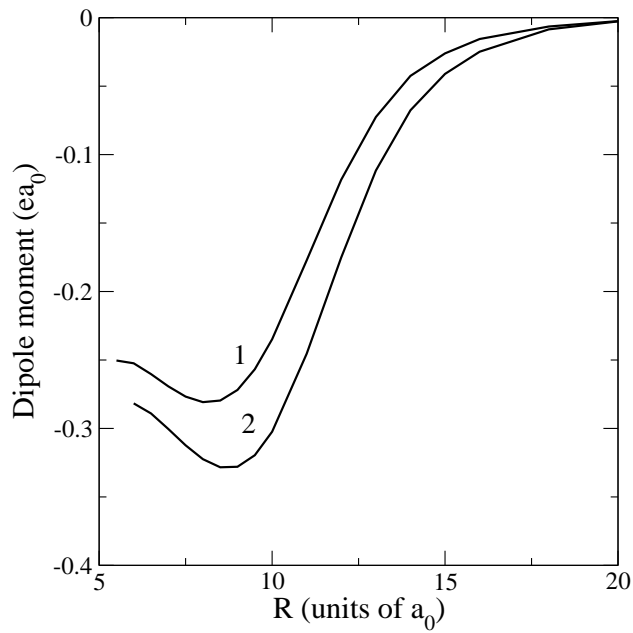}
\caption{The electric dipole moment of the $X^1\Sigma^+$ ground state of the
KRb dimer as a function of internuclear separation. The dipole moment is found
from a nonrelativistic (curve 1) and a relativistic (curve 2) calculation.}
\label{1s_dipole}
\end{figure}

\begin{figure} \vspace*{12.5cm} \includegraphics{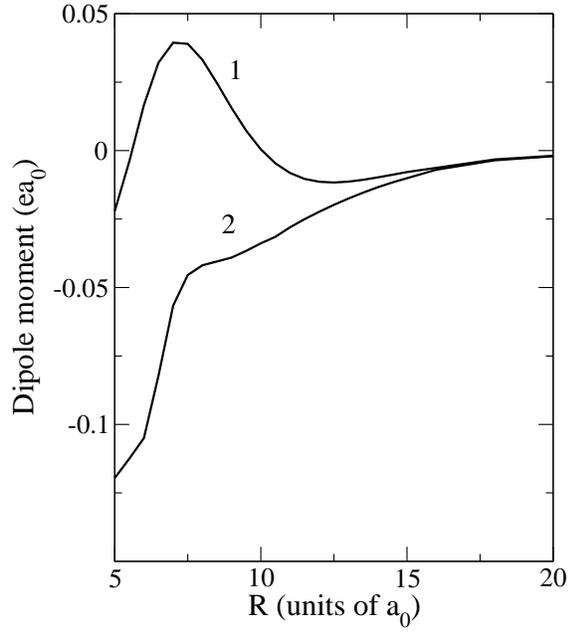}
\caption{The electric dipole moment of the $a^3\Sigma^+$ $\Omega$ =
0 state of KRb as a function of internuclear separation. The
dipole moment is found from a nonrelativistic (curve 1) and relativistic (curve 2) calculation.} \label{3s_dipole}
\end{figure}

Figures \ref{1s_dipole} and \ref{3s_dipole} show the electric dipole
moments of the $X^1\Sigma^+$ and $a^3\Sigma^+$ potentials of KRb
as a function of internuclear separation, respectively.  A negative
dipole moment implies an excess electron charge on the potassium atom.
The presented dipole moments are calculated with the relativistic
and nonrelativistic Hamiltonian. Curve 1 in Figs. \ref{1s_dipole} and
\ref{3s_dipole} is determined from the most accurate nonrelativistic
calculation, which includes the two-electron occupation of all
orbitals upto $6p$ for the K atom and $7p$ or the Rb atom.  Curve 2
in Figs. \ref{1s_dipole} and \ref{3s_dipole} shows the $\Omega=0$
relativistic calculation. The number of excited orbitals is smaller than
in the nonrelativistic basis set and limited to excitations upto $4d$
and $5p$ for K atom and $5d$ and $6p$ for Rb atom.  The difference between
the dipole moments of the $\Omega=0$ and 1 component of the $a^3\Sigma^+$
potential is much smaller than the uncertainties of our calculations.
Tables \ref{tab1} and \ref{tab2} tabulate the permanent dipole moment
of the $X^1\Sigma^+$ and $a^3\Sigma^+$ potentials shown in the figures.

Our calculation shows that the distribution of the charge density between
K and Rb is very diffuse. It means that the dipole moment depends not only
on the electron charge transfer from Rb to K atom but also on the induced
polarization from the charge transfer, i.e. the electrons occupy excited
$p$ and $d$ orbitals.  Charge transfer and the induced polarization are
of opposite sign.  This is especially true for the $a^3\Sigma$ state.
We find that the excess charge is almost equal for the singlet and
triplet states, but the latter has a smaller dipole moment.

\begin{figure} \vspace*{12.1cm} \includegraphics{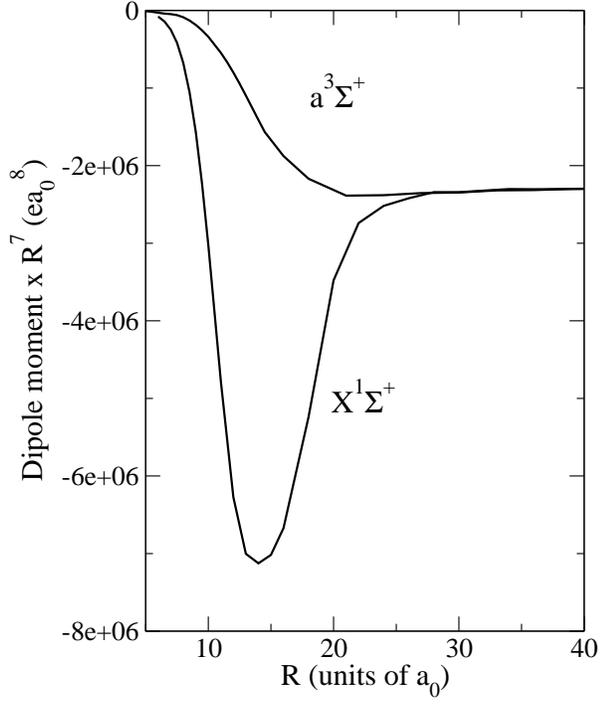}
\caption{The relativistic electric dipole moment multiplied by $R^7$
of the $X^1\Sigma^+$ and $a^3\Sigma^+(\Omega=0)$ states of KRb as
a function of internuclear separation.} \label{R7}
\end{figure}

The permanent dipole moments of the $X^1\Sigma^+$ and $a^3\Sigma^+$
state multiplied by $R^7$, based on a the relativistic calculation are
shown in Fig.~\ref{R7}. It shows that the long-range behavior of the
dipole moment of the two states are equal and proportional to $1/R^7$.
This long-range behavior is due to the modification of the molecular
wave function by the dipole-dipole and dipole-quadrupole multipole
interactions \cite{Whisnant}.

We believe that our convergence with respect to correlation or the
number of basis function is of the same order as the difference between
nonrelativistic and relativistic calculation.
Consequently, we feel that the most accurate dipole moment is obtained
from averaging the nonrelativistic and relativistic calculation and
assuming an one-standard deviation uncertainty equal to half the
difference between the two calculations. For the $X^1\Sigma^+$ state
the dipole moment is $-$0.30(2) $ea_0$ at $R_e=7.7$ $a_0$, while for
the $a^3\Sigma^+$ state the dipole moment is $-$0.02(1) $ea_0$ at
$R_e$=11.2 $a_0$.

\section{Transition dipole moments} \label{tdm}

In this section we analyze the possibility of the creation of dipolar
KRb molecules in an optical lattice. We assume that colliding atoms
are initially in the doubly spin-polarized state, which only allow
them to come together on the state $a^{3}\Sigma^+$ potential. Thus
photoassociation via an excited molecular $^{3}\Sigma$ or $^{3}\Pi$
state produces $a^{3}\Sigma^+$ molecular levels. In principle this makes
the production of molecules in the $X^{1}\Sigma^+$ state problematical,
since one has to get from the triplet to singlet spin manifold. There is,
however, a viable route from the doubly spin-polarized colliding ground
state atoms to ground $X^{1}\Sigma$ levels via the excited state. When
the detuning of the photoassociation laser from atomic resonance is
small compared to the excited state spin-orbit splitting, as will be
the case for most practical photoassociation schemes, then an excited
$\Omega$ = 0 or 1 Hund's case (c) state can have both singlet and triplet
character. Thus, an excited molecular state can be formed from excitation
from a $a^{3}\Sigma^+$ state that can re-emit light in a transition to
a $X^{1}\Sigma^+$ state. This process is absent in homonuclear dimers,
since the additional gerade- ungerade selection rule for electronic
transitions prevents it.

\begin{figure} \vspace*{12.5cm} \includegraphics{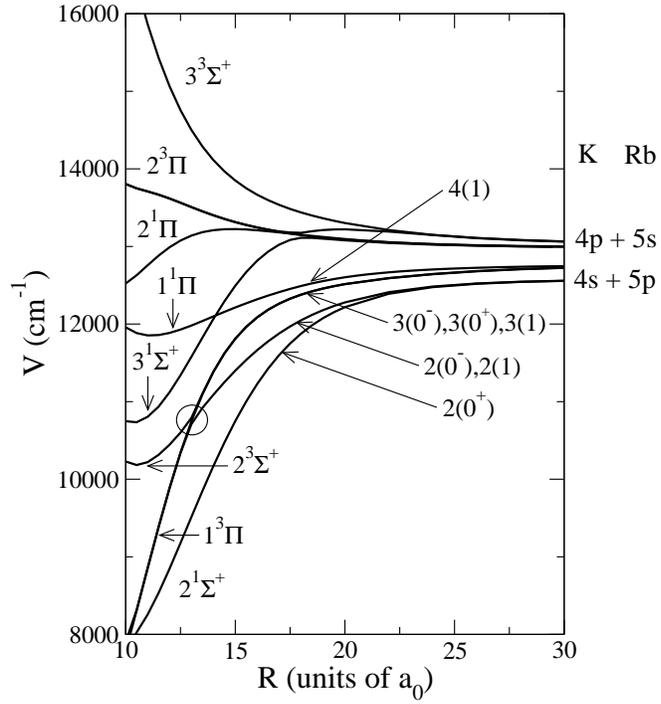}
\caption{Electronic $\Omega$ = 0$^\pm$, 1 potentials of excited states of
the KRb dimer in an intermediate region of internuclear separation. The
short-range $n^{(2S+1)}\Lambda^\pm$ and long-range $m(\Omega^\pm)$
labeling is indicated.  The numbers $n$ and $m$ show the energy-ordered
appearance of these states.}
\label{excpot} 
\end{figure}

Figure~\ref{excpot} shows the $\Omega=0^\pm$ and $1$ relativistic excited
state potentials as a function of internuclear separation.  The excited
state potentials are obtained with the basis described in Sec.~\ref{gs}.
At short internuclear separation the potentials are described by the
$^{(2S+1)}\Lambda^\pm$ Hund's case (a) labeling following Ref.~\cite{Park}
where $\Lambda$ is a  projection of the electronic orbital angular momentum
on the molecular axis and $S$ is the total electron spin.
At longer $R$ the curves are better described by $\Omega^\pm$ Hund's case
(c) labeling and relativistic effects are important.  The potentials
dissociate to the excited K(4s)+Rb(5p $^2P_j$) and K(4p $^2P_j$ )+Rb(5s)
fine structure limits. The two lowest dissociation limits correspond to
the two fine structure states of the excited Rb atom plus a ground state K
atom. The long-range behavior of the relativistic potentials dissociating
to these two limits is attractive.  There are seven attractive potentials
with $\Omega=0$ and $1$.
Our potentials agree with the calculations of Refs.~\cite{Park,Rousseau}.

\begin{figure} \vspace*{12.1cm} \includegraphics{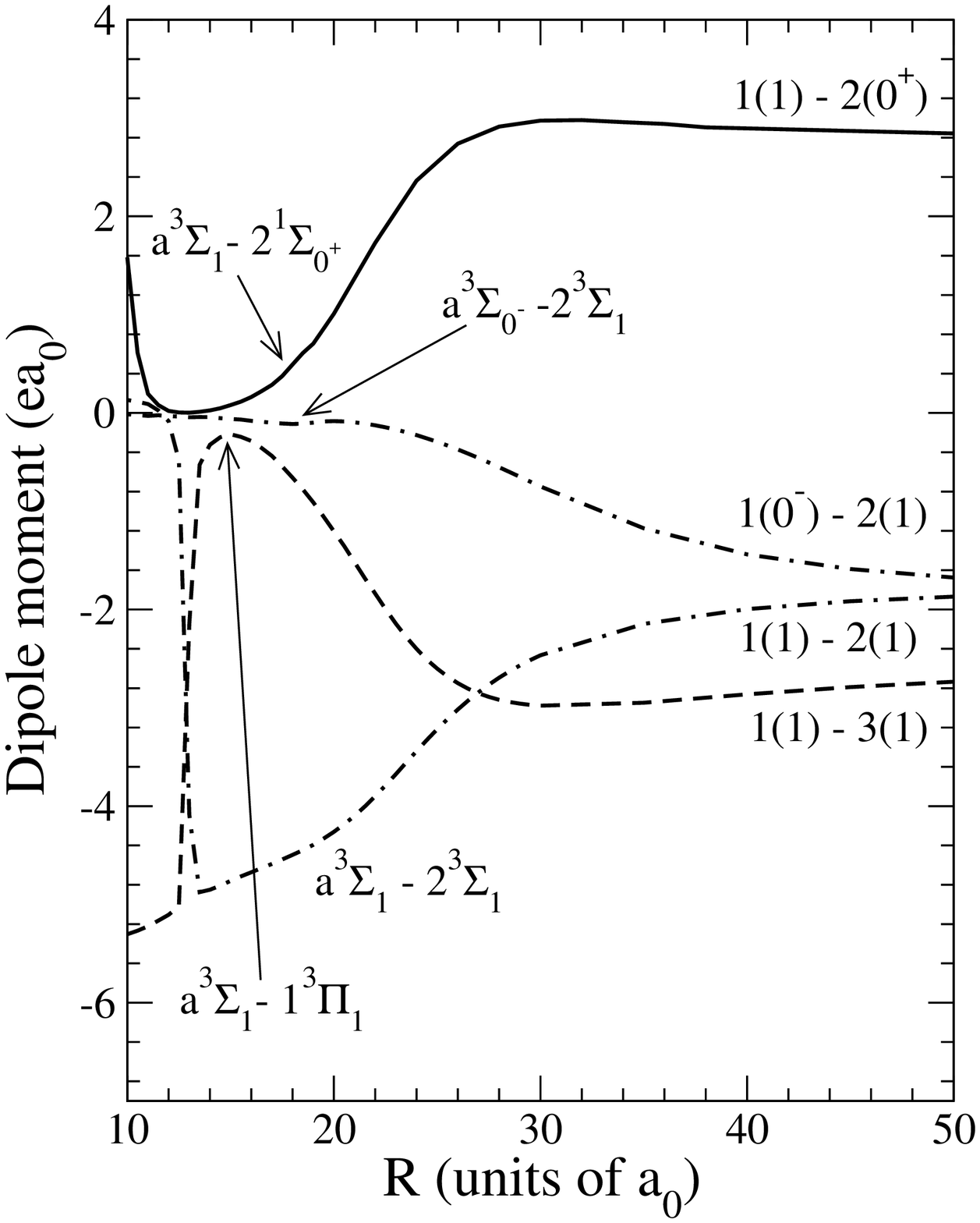}
\caption{Transition dipole moments between the ground $a^3\Sigma^+$ and excited
states of the KRb dimer as a function of internuclear separation. 
The curves are labeled by short- and long-range symmetries as in 
Fig.~\ref{excpot}}
\label{3s_trans}
\end{figure}

\begin{figure} \vspace*{12.1cm} \includegraphics{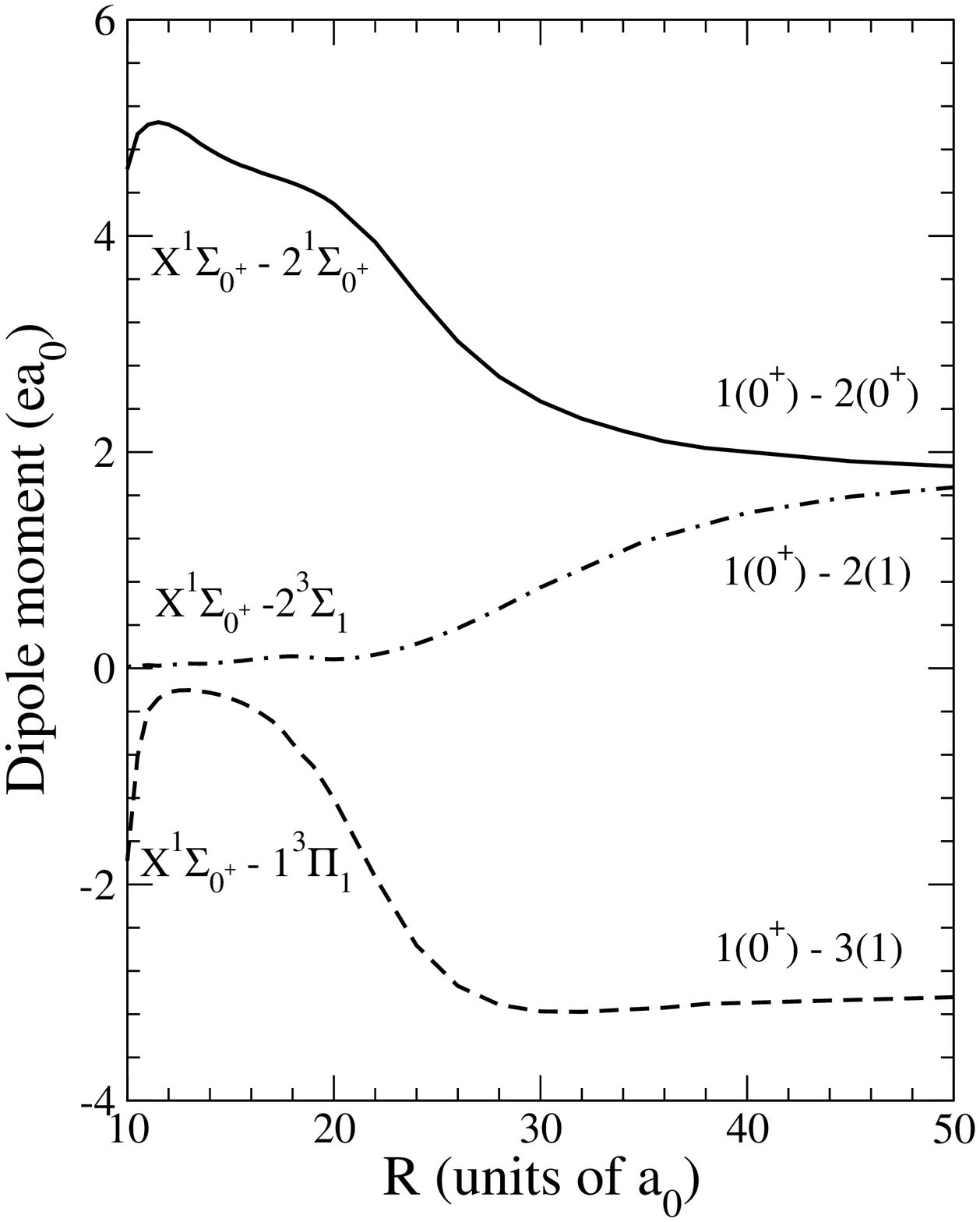}
\caption{Transition dipole moments between the ground $X^1\Sigma^+$ and excited
states of the KRb dimer as a function of internuclear separation. 
The curves are labeled by short- and long-range symmetries as in
Fig.~\ref{excpot}}
\label{1s_trans}
\end{figure}

The attractive excited-state potentials are the most likely
candidates for use in two-color Raman photoassociation experiments.
We calculate transition dipole moments relevant for transitions from
the $a^3\Sigma^+(0^-,1)$ state, through the attractive $\Omega=0^\pm$
and 1 potentials, to the $X^1\Sigma^+(0^+)$ state. Figures~\ref{3s_trans}
and \ref{1s_trans} show the nonzero transition dipole moments from the
$a^3\Sigma^+$ and $X^1\Sigma^+$ states, respectively. Curves with the same
style in the two figures indicate the same intermediate excited state.
Photon selection rules ensure that $0^+$ $\to$ $0^-$ transition
are not allowed. At long-range the transition dipole moments become
independent of $R$ and their absolute values approach the Rb 5s $\to$
5p($^2P_{j}$) transition dipole moment when the corresponding excited
potential dissociates to the K(4s)+Rb(5p $^2P_j$) limit.

At short-range internuclear separation the dipole moments strongly
depend on $R$. This behavior reflects the change from a Hund's case (a)
to a relativistic coupling scheme between 20 $a_0$ and 30 $a_0$.
In Fig.~\ref{3s_trans} the $a^3\Sigma^+(\Omega=1)$ to $2^1\Sigma^+(0^+)$
and in Fig.~\ref{1s_trans} the $X^1\Sigma^+(0^+)$ to $2^3\Sigma^+(1)$
transition dipole moments approach zero at short $R$ because singlet
to triplet transitions are not allowed. In  Fig.~\ref{3s_trans} both
$a^3\Sigma^+(1)$ to $2^3\Sigma^+(1)$ and $a^3\Sigma^+(0^-)$ to
$2^3\Sigma^+(1)$ have a small dipole moment at small internuclear
separation. This can be understood by noting that the structure of the
ground and excited state potentials shown in Figs.~\ref{potentials}
and \ref{excpot} is similar to that of homonuclear alkali-metal dimers,
where gerade/ungerade symmetry is valid.  In homonuclear dimers electric
dipole transitions between gerade and ungerade states are forbidden.

The sudden change in dipole moment in Fig.~\ref{1s_trans} near 13 $a_0$
is related to the avoided crossing indicated by the circle in Fig.~\ref{excpot} 
between the $\Omega=1$ components of
1$^3\Pi$ and 2$^3\Sigma^+$ potentials.  We estimate that the uncertainty
of the transition dipole moments is 0.1 $ea_0$ based on a comparison of the
calculated dipole moments at $R=100\,a_0$ with the known 5s to 5p dipole
moments of Rb\cite{nist}.

\section{Conclusion}

We determined the permanent dipole moments of the $^{1,3}\Sigma^+$
states of the ground configuration of the KRb heteronuclear molecule
using a nonrelativistic and relativistic CI valence bond method.  The KRb
permanent dipole moments are small compared to ``truly'' polar molecules,
such as NaCl. It might, however, be large enough for experiments that aim to
confine KRb in optical traps.  In addition we calculated the potential
energy curves and transition dipole moments to excited states correlating
to K(4s)+Rb(5p) atomic limits.  We have shown that there exist allowed
transitions starting from colliding doubly polarized K and Rb atoms via
an exited state to the singlet X$^{1}\Sigma^+$ state.

The calculation of the electric dipole moments is a first step towards obtaining
quantitative estimates of photoabsorption and molecular production rates in
a gas of K and Rb atoms. In the future we plan to evaluate Frank-Condon
factors between vibrational levels of ground and excited state
potentials. In addition we need to investigate the effect of black-body
radiation on the dipolar molecule by evaluating Frank-Condon
factors between vibrational levels of the ground state.

\section{Acknowledgments}

We wish to acknowledge helpful discussions with Ilia Tupitsyn and Andrea Simoni.

\begin{table} \caption{
The electric dipole moment of the $X^1\Sigma^+$ state. Both nonrelativistic
($nrel$) and relativistic ($rel$) results are presented. The internuclear
separation $R$ is in units of $a_0$  and the dipole moment is in units
of $ea_0$.}  \label{tab1}
\begin{center} 
\begin{tabular}{lcc} 
$R$    &  $nrel$    & $rel$ \\ \hline 
6  &  -2.52$\times$ 10$^{-1}$  &  -2.82$\times$ 10$^{-1}$  \\ 
6.5  &  -2.60$\times$ 10$^{-1}$  & -2.89$\times$ 10$^{-1}$  \\ 
7  &  -2.69$\times$ 10$^{-1}$  &  -3.00$\times$ 10$^{-1}$  \\ 
7.5  &  -2.77$\times$ 10$^{-1}$  & -3.12$\times$ 10$^{-1}$  \\ 
8  &  -2.81$\times$ 10$^{-1}$  &  -3.22$\times$ 10$^{-1}$  \\ 
8.5  &  -2.80$\times$ 10$^{-1}$  & -3.28$\times$ 10$^{-1}$  \\ 
9  &  -2.72$\times$ 10$^{-1}$  &  -3.28$\times$ 10$^{-1}$  \\ 
9.5  &  -2.57$\times$ 10$^{-1}$  & -3.20$\times$ 10$^{-1}$  \\ 
10  &  -2.35$\times$ 10$^{-1}$  &  -3.02$\times$ 10$^{-1}$  \\ 
11  &  -1.77$\times$ 10$^{-1}$  & -2.45$\times$ 10$^{-1}$  \\ 
12  &  -1.18$\times$ 10$^{-1}$  &  -1.75$\times$ 10$^{-1}$  \\ 
13  &  -7.25$\times$ 10$^{-2}$  & -1.13$\times$ 10$^{-1}$  \\ 
14  &  -4.25$\times$ 10$^{-2}$  &  -6.80$\times$ 10$^{-2}$  \\ 
15  &  -2.61$\times$ 10$^{-2}$  & -4.11$\times$ 10$^{-2}$  \\ 
16  &  -1.56$\times$ 10$^{-2}$  &  -2.36$\times$ 10$^{-2}$  \\ 
18  &  -6.44$\times$ 10$^{-3}$  & -9.44$\times$ 10$^{-3}$  \\ 
20  &  -2.48$\times$ 10$^{-3}$  &  -2.48$\times$ 10$^{-3}$  \\ 
22  &  -1.10$\times$ 10$^{-3}$  & -1.10$\times$ 10$^{-3}$  \\ 
24  &  -5.49$\times$ 10$^{-4}$  &  -5.49$\times$ 10$^{-4}$  \\ 
26  &  -3.01$\times$ 10$^{-4}$  & -3.01$\times$ 10$^{-4}$  \\ 
28  &  -1.74$\times$ 10$^{-4}$  &  -1.74$\times$ 10$^{-4}$  \\ 
30  &  -1.08$\times$ 10$^{-4}$  & -1.08$\times$ 10$^{-4}$  \\ 
32  &  -6.88$\times$ 10$^{-5}$  &  -6.88$\times$ 10$^{-5}$  \\ 
34  &  -4.57$\times$ 10$^{-5}$  & -4.57$\times$ 10$^{-5}$  \\ 
36  &  -3.02$\times$ 10$^{-5}$  &  -3.02$\times$ 10$^{-5}$  \\ 
38  &  -2.06$\times$ 10$^{-5}$ &  -2.06$\times$ 10$^{-5}$  \\ 
40  &  -1.43$\times$ 10$^{-5}$  &  -1.43$\times$ 10$^{-5}$  \\ 
\end{tabular}
\end{center}
\end{table}

\begin{table}
\caption{ Electric dipole moment of the $a^3\Sigma^+$ state. Both
nonrelativistic ($nrel$) and relativistic ($rel$) $\Omega$ =0 results
are presented.} \label{tab2}
\begin{center}
\begin{tabular}{lcc}
$R$  &   nrel     &     rel \\ \hline
5  &  -2.22$\times$ 10$^{-2}$  &  -1.20$\times$ 10$^{-1}$  \\
5.5  &  -3.94$\times$ 10$^{-3}$  &  -1.13$\times$ 10$^{-1}$  \\
6  &  1.66$\times$ 10$^{-2}$  &  -1.05$\times$ 10$^{-1}$  \\
6.5  &  3.21$\times$ 10$^{-2}$  &  -8.21$\times$ 10$^{-2}$  \\
7  &  3.94$\times$ 10$^{-2}$  &  -5.66$\times$ 10$^{-2}$  \\
7.5  &  3.90$\times$ 10$^{-2}$  &  -4.54$\times$ 10$^{-2}$  \\
8  &  3.32$\times$ 10$^{-2}$  &  -4.19$\times$ 10$^{-2}$  \\
8.5  &  2.46$\times$ 10$^{-2}$  &  -4.05$\times$ 10$^{-2}$  \\
9  &  1.56$\times$ 10$^{-2}$  &  -3.91$\times$ 10$^{-2}$  \\
9.5  &  7.26$\times$ 10$^{-3}$  &  -3.71$\times$ 10$^{-2}$  \\
10  &  4.50$\times$ 10$^{-4}$  &  -3.45$\times$ 10$^{-2}$  \\
10.5  &  -4.67$\times$ 10$^{-3}$  &  -3.16$\times$ 10$^{-2}$  \\
11  &  -8.18$\times$ 10$^{-3}$  &  -2.86$\times$ 10$^{-2}$  \\
11.5  &  -1.03$\times$ 10$^{-2}$  &  -2.56$\times$ 10$^{-2}$  \\
12  &  -1.14$\times$ 10$^{-2}$  &  -2.28$\times$ 10$^{-2}$  \\
12.5  &  -1.17$\times$ 10$^{-2}$  &  -2.01$\times$ 10$^{-2}$  \\
13  &  -1.13$\times$ 10$^{-2}$  &  -1.77$\times$ 10$^{-2}$  \\
13.5  &  -1.07$\times$ 10$^{-2}$  &  -1.54$\times$ 10$^{-2}$  \\
14  &  -9.80$\times$ 10$^{-3}$  &  -1.34$\times$ 10$^{-2}$  \\
14.5  &  -8.82$\times$ 10$^{-3}$  &  -1.16$\times$ 10$^{-2}$  \\
15  &  -7.83$\times$ 10$^{-3}$  &  -9.02$\times$ 10$^{-3}$  \\
18  &  -3.24$\times$ 10$^{-3}$  &  -3.24$\times$ 10$^{-3}$  \\
21  &  -1.22$\times$ 10$^{-3}$  &  -1.22$\times$ 10$^{-3}$  \\
24  &  -4.78$\times$ 10$^{-4}$  &  -4.78$\times$ 10$^{-4}$  \\
27  &  -2.04$\times$ 10$^{-4}$  &  -2.04$\times$ 10$^{-4}$  \\
30  &  -9.67$\times$ 10$^{-5}$  &  -9.67$\times$ 10$^{-5}$  \\
33  &  -5.18$\times$ 10$^{-5}$  &  -5.18$\times$ 10$^{-5}$  \\
36  &  -2.72$\times$ 10$^{-5}$  &  -2.72$\times$ 10$^{-5}$  \\
39  &  -1.54$\times$ 10$^{-5}$  &  -1.54$\times$ 10$^{-5}$  \\
42  &  -9.14$\times$ 10$^{-6}$  &  -9.14$\times$ 10$^{-6}$  \\
\end{tabular}
\end{center}
\end{table}

\begin{references}

\bibitem{You} S. Yi and L. You, Phys. Rev. A {\bf 61}, 041604 (2000).

\bibitem{Goral}  K. G\'oral, K. Rzaszewski, and T. Pfau,
 Phys. Rev. A  {\bf 61}, 051601 (2000).

\bibitem{Santos} L. Santos {\it et al}., Phys. Rev. Lett. {\bf 85}, 1791 (2000).

\bibitem{Goral01} K. G\'oral and L. Santos, cond-mat/0203542.

\bibitem{Meystre}
H. Pu, W. Zhang, and P. Meystre, Phys. Rev. Lett. {\bf 87}, 140405 (2001).

\bibitem{Giovanazzi}
S. Giovanazzi, D.O'Dell, and G. Kurizki, Phys. Rev. Lett. {\bf 88}, 130402
(2002)

\bibitem{Jaksch}D. Jaksch, V. Venturi, J. I. Cirac, C. J. Williams, and
P. Zoller, Phys. Rev. Lett. {\bf 89}, 040402 (2002).

\bibitem{Damski}B. Damski, L. Santos, E. Tiemann, M. Lewenstein, S. Kotochigova,
P. Julienne, and P. Zoller, Phys. Rev. Lett. to be published, (2003).

\bibitem{Moore}M. G. Moore, H. R. Sadeghpour, cond-mat/0209621.

\bibitem{Kotochigova} S. Kotochigova, E. Tiesinga, and I. Tupitsyn,
in "New Trends in Quantum Systems in Chemistry and Physics",
{\bf 1}, 219 (Kluwer Academic Publ., The Netherlands, 2001).

\bibitem{Ross}A. J. Ross, C. Effantin, P. Crozet, and E. Boursey, J. Phys. B
{\bf 23}, L247 (1990).

\bibitem{Okada}N. Okada, S. Kasahara, T. Ebi, M. Baba, and H. Kato, J. Chem.
Phys. {\bf 105}, 3458 (1996).

\bibitem{Kasahara}S. Kasahara, C. Fujiwara, N. Okada, H. Kato, and
M. Baba, J. Chem.  Phys. {\bf 111}, 8857 (1999).

\bibitem{Amiot}C. Amiot, and J. Verges, J. Chem. Phys. {\bf 112}, 7068 (2000).

\bibitem{Yiannopoulou}A. Yiannopoulou, T. Leininger, A. M. Lyyra, and G.-H. Jeung,
Int. J. Quant. Chem. {\bf 57}, 575 (1996).

\bibitem{Leininger}T. Leininger, H. Stoll, and G.-H. Jeung, J. Chem. Phys.
{\bf 106}, 2541 (1997).

\bibitem{Park}S. J. Park, Y. J. Choi, Y. S. Lee, and G.-H. Jeung,
Chem. Phys. {\bf 257}, 135 (2000).

\bibitem{Rousseau}S. Rousseau, A. R. Allouche, and M. Aubert-Frecon, J. Mol.
Spectrosc. {\bf 203}, 235 (2000).

\bibitem{Bussery}B. Bussery, Y. Ackhar, and M. Aubert-Frecon, Chem. Phys. {\bf 116},
319 (1987).

\bibitem{Patil}S. H. Patil and K. T. Tang, J. Chem. Phys. {\bf 106}, 2298 (1996).

\bibitem{Wang}H. Wang and C. Stwalley, J. Chem. Phys. {\bf 108}, 5767 (1998).

\bibitem{Marinescu}M. Marinescu and H. R. Sadeghpour, Phys. Rev. A {\bf 59}, 390 
(1990).

\bibitem{Derevianko}A. Derevianko, J. F. Babb, and A. Dalgarno, Phys. Rev. A
{\bf 63} 052704 (2001).

\bibitem{Whisnant}D. M. Whisnant and W. Byers Brown, Mol. Phys. {\bf 26}, 1105 (1973).

\bibitem{nist}http://physics.nist.gov/PhysRefData/contents-atomic.html

\end{references}
\end{document}